\documentclass[twocolumn,showpacs,preprintnumbers,amsmath,amssymb]{revtex4}
\usepackage{graphicx}

\begin{document}

\title{Model of Multiphoton Transitions in a Current-Biased Josephson Junction}

\author{I. Tornes} \email{itornes@mps.ohio-state.edu} 

\author{D. Stroud} \email{stroud@mps.ohio-state.edu}

\affiliation{Department of Physics, The Ohio State University,Columbus, Ohio 43210}

\begin{abstract}
We present a simple model to describe multiphoton transitions
between the quasi bound states of a current-driven Josephson
junction.  The transitions are induced by applying an ac voltage
with controllable frequency and amplitude across the junction.  The
voltage induces transitions across the junction when the frequency
$\omega$ satisfies $n\hbar\omega = \Delta E_{10}$, where 
$\Delta E_{10}$ is the splitting between the ground and first excited
quasi-bound state of the junction.  We calculate the matrix elements
of the transitions as a function of the dc bias current $I$, and
the frequency $\omega$ and amplitude $V_{ac}$ of the microwave voltage, for representative
junction parameters.  We also calculate the frequency-dependent absorption
coefficient by solving the relevant Bloch equations when the ac voltage
is sufficiently weak.  In this regime, the absorption coefficient is
a sum of Lorentzian lines centered at the n-photon absorption frequency,
of strength proportional to the squared matrix elements.  For
fixed $V_{ac}$, the transition rate for an $n$-photon process
usually decreases with increasing $n$.  We also find a characteristic
even-odd effect: the absorption coefficient typically increases with $I$
for $n$ even but decreases for $n$ odd.  Our results agree qualitatively
with recent experiments.
%IVAN2: should I say that the results agree qualitatively with recent
%experiments?
%DAVID2:  I think using qualitatively is fine.  They have the same behavior, but a
%direct quantitative comparison cannot be made.  

%IVAN, I just guessed at the last sentence - is it actually true?
%DAVID, Yes, I think this is right based on the results that I obtained

\end{abstract}

\date{\today}

\maketitle

\section{Introduction}

A current-driven Josephson junction can exhibit clear experimental
manifestations of quantized energy levels\cite{martinis1,clarke,martinis2,cooper,
makhlin, silvestrini,yu}.  
%removed nakamura, chen, and tsai because this is a charge qubit.  
For example, if the
junction is biased so that the first minimum of the washboard
potential holds only a few quasi-bound states, a suitable ac current
can induce transitions between the lowest levels.  Population of the
excited levels leads to tunneling through the barrier and hence a
voltage pulse which can readily be detected experimentally.  In part
because of this behavior, such junctions are promising candidates
for use as qubits (``phase qubits'') in quantum computation\cite{martinis}.

\begin{figure}
\begin{center}
\includegraphics[angle = 0, width = 0.5\textwidth]{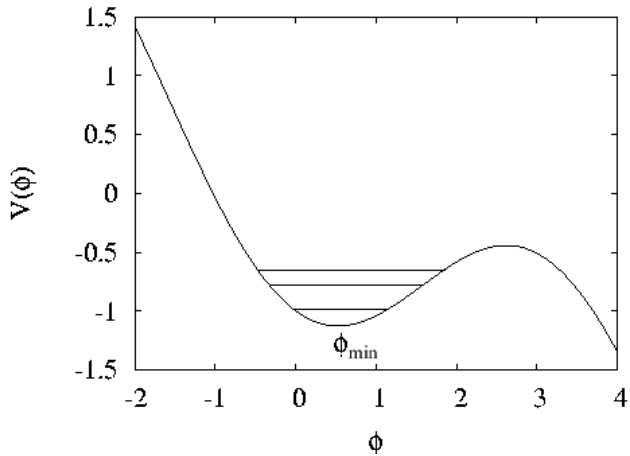}
\caption
{ \small Schematic diagram of the level structure for a Josephson junction subjected to an
applied dc current.}\label{washboard}
\end{center}
\end{figure}

Fig.~\ref{washboard} shows a schematic of the level structure for such a
junction.  Since the potential near a minimum is
anharmonic, the level spacings are unequal, as indicated in the
sketch. If $\Delta E_{10} = E_1 - E_0$ is the spacing between the
ground and first excited state, and $\Delta E_{21} = E_2 - E_1$ is
that between the first and second excited level, then $\Delta E_{21}
< \Delta E_{10}$. An ac perturbation of frequency $\omega = \Delta
E_{10}/\hbar$ will induce transitions between the lowest two states, but
because of the anharmonicity, will not produce further excitations
to the next level.  Rabi oscillations between the lowest
two levels have been demonstrated experimentally - that is, the population of the
excited level oscillates with a frequency related to the {\em
amplitude} of the ac pulse, if that perturbation is tuned to be in
resonance with $\Delta E_{10}$\cite{martinis}.

Recently, Wallraff {\it et al}\cite{wallraff2} have demonstrated
{\em multiphoton} transitions between the ground and first excited
states of a current-biased Josephson junction.  In this experiment,
transitions were induced between the two levels by subjecting the
system to an ac voltage at a frequency $\Delta E_{10}/(n\hbar)$,
with $n > 1$.  Similar transitions have also been observed in flux
qubits by Saito {\it et al}\cite{saito}, and in general in qubits
based on anharmonic potentials containing more than two quasi-bound
states\cite{dykman}.   These transitions correspond to absorption of
n microwave photons. Such transitions can be used to generate
photons at multiples of the incident frequency.  If the states
involved have a sufficiently long coherence time, the perturbation
could also lead to Rabi oscillations arising from multiphoton
transitions.  Indeed, such Rabi oscillations have been reported
experimentally\cite{saito1}.  Thus, they could be a significant advance in using
these phase qubits in quantum computation.

A number of workers have discussed models for Josephson junctions driven
by strong ac perturbations.  For example, Ashhab {\it et al.}\cite{ashhab} 
have considered a two-level system subject to a strong perturbation which is
harmonic in time, and demonstrate the occurrence of resonances at frequencies
$\omega = \Omega/n$, where $\omega$ is the drive frequency, $\hbar\Omega$ is the unperturbed
level splitting, and $n$ is a positive integer.  Saito {\it et al.}\cite{saito1} have
observed and analyzed Rabi oscillations corresponding to n-photon transitions in
a superconducting flux qubit.  Inomata {\it et al}\cite{inomata} have measured and analyzed
macroscopic quantum tunneling in intrinsic Bi$_2$Sr$_2$CaCu$_2$O$_{8+\delta}$ Josephson 
junctions arising from multiphoton transitions, using an analysis based on the classical
Josephson junction equation of motion.  Koval {\it et al.}\cite{koval} have analyzed the
enhancement of Josephson phase diffusion by microwaves, starting from
the classical equation for a resistively and capacitively shunted
Josephson junction.  They obtain a characteristic dependence
of the n-photon absorption coeficient on the square of an n$^{th}$ order Bessel function, similar
to that obtained by earlier workers\cite{odintsov,falci}, and also resembling that found in the different analysis presented here.   Fistul {\it et al.}\cite{fistul} have analyzed the 
quantum escape of the phase in a strongly driven Josephson junction in the presence of both
ac and dc bias curents, in good agreement with their own experiments.   Goorden and Wilhelm\cite{goorden} analyzed nonlinear driving effects in a continuously driven solid-state qubit using the Bloch-Redfield
equation, once again obtaining n-photon resonance effects.  Gr\/{o}nbech-Jensen {\it et al.}\cite{jensen}
have obtained multiphoton-like effects in Josephson junctions computationally, by solving the classical equation
of motion for a junction in the presence of both dc and ac currents.  A number of other workers have considered
effects of strong ac fields on Josephson junctions in various geometries, mostly in the context of their
possible use as qubits for quantum computation\cite{strauch,berkley,shnyrkov,ivlev}.

In this paper, we provide a simple model to describe this
multiphoton absorption in Josephson phase qubits, using a somewhat different
approach from those described above.  In our model, we
consider a Josephson junction in the presence of a dc driving
current plus an ac voltage.  The junction is assumed to have very
little dissipation, so that the only terms in the Hamiltonian are
the Josephson coupling and a capacitive energy.   These two combine
to produce the well-known pendulum-like Hamiltonian of the junction.
The spacing between the levels of this system depends on the current
bias.

When the ac voltage is introduced, the gauge-invariant phase difference in the
Hamiltonian includes a term arising from that voltage.  Use of a Bessel 
function expansion then shows that
this extra term produces $n$-photon transitions between the ground
and first excited state of the current-biased junction.  We
calculate the $n$-photon transition rate as a function of ac voltage amplitude
and dc driving current, using the Fermi golden
rule.  Our model readily produces the transitions seen in
experiments.

The remainder of this paper is organized as follows. In the next
section, we briefly describe our calculation of the energy
eigenvalues and wave functions of the time-independent Hamiltonian.
Following this, we describe the expansion which leads to all the
n-photon terms in the Hamiltonian.  Finally, we present numerical
results for the n-photon transition rate, using the Fermi golden
rule.  A brief discussion follows in Section V.

\section{Formalism}

\subsection{Current-Biased Josephson Junction}

\subsubsection{Hamiltonian}

 In the absence of resistive shunting, a
capacitively shunted Josephson junction driven by a current $I$ can
be described by the Hamiltonian\cite{tinkham}
$H = (2en)^2/(2C) -\frac{\hbar I_c}{2e}\cos\phi - \frac{\hbar I}{2e}\phi$,
where $I_c$ and $I$ are the junction critical current and the
applied current, $\phi$ is the phase difference across the junction, $C$ is
the junction capacitance, $n$ is the Cooper pair number difference
across the junction, and $2e$ is the charge of a Cooper pair.
$n$ and $\phi$ are canonically conjugate operators,
satisfying the commutation relation $[n, \phi] = -i$.  This relation
can be satisfied if we use the representation $n = -
i\partial/\partial\phi$ for the number operator.  In this
representation, $H$ becomes
\begin{equation}
H = -E_C\frac{\partial^2}{\partial\phi^2}-\frac{\hbar I_c}{2e}\cos\phi -\frac{\hbar I}{2e}\phi
\equiv -E_C\frac{\partial^2}{\partial\phi^2} + U(\phi),
\label{eq:hwash1}
\end{equation}
where $E_C = 2e^2/C$ is the charging energy of the junction.  Eq.\
(\ref{eq:hwash1}) describes a phase "particle" moving in a
"washboard potential,"
$U(\phi) = -\frac{\hbar I_c}{2e}\cos\phi - \frac{\hbar I}{2e}\phi$,
whose slope is controlled by $I$.
In the experiments of Ref.~\cite{martinis}, $E_C \ll E_J \equiv
\hbar I_c/(2e)$.  Although this regime corresponds to a relatively
large junction, it must still be treated quantum-mechanically when
the dissipation is very small.  For their parameters, the observed quasi-bound state transitions
occur when $I/I_c \sim 0.99$.  In our calculations, for calculational convenience,
we have used a larger ratio of $E_C/E_J$.  
%IVAN2: is this comment consistent with other remarks about this ratio?
%DAVID2:  Yes it is.
In order to model the relevant
experiments, we solve both the time-independent and the
time-dependent Schr\"{o}dinger equations for the above Hamiltonian, as we now describe.

\subsubsection{Solution of the Time-Independent Schr\"{o}dinger
Equation}

 For $I < I_c$, $U(\phi)$ has
local potential minima at $\phi = \phi_k \equiv \sin^{-1}(I/I_c) +
2\pi k$, $k = 0$, $\pm 1$, $\pm 2$,.... For a given $I$, the
potential near the first of these minima varies as $U(\phi) \sim
U_{min} + K(I)(\phi-\phi_0)^2/2$, where $U_{min} = U(\phi_0)$ and
$K(I)$ is a current-dependent second derivative.  If the potential
were {\em strictly} harmonic near the minimum, the solutions of the
time-independent Schr\"{o}dinger equation would be harmonic
oscillator eigenstates, with energies $E_n(I) =
(n+\frac{1}{2})\hbar\omega_c(I)$.  These
frequencies are readily shown to satisfy
\begin{equation}
\omega_c(I) = \omega_c(0)\left[1 - \frac{I^2}{I_c^2}\right]^{1/4},
\label{eq:omegac}
\end{equation}
where $\omega_c(0) = [2e I_c/(\hbar C)]^{1/2}$ is the harmonic
oscillator frequency at $I = 0$, usually called the junction plasma
frequency.  Because the potential is actually anharmonic, there are
two corrections to the above expression for the energies of the
eigenstates: the junction levels are not equally spaced as in a
harmonic potential, and they are only quasi-bound states,
because they can tunnel out of the well.

It is convenient to work with a scaled Hamiltonian
\begin{equation}\label{eq:hscale}
\tilde{H} = \frac{H}{E_J} = -\cos\phi - \frac{I}{I_c}\phi -
\frac{E_C}{E_J}\frac{\partial^2}{\partial\phi^2}.
\end{equation}
The parameters appropriate for the experiments of Ref.\
\cite{martinis} are $E_J \sim 6.8\times 10^{-14}$ ergs and $E_C\sim
8.3\times 10^{-20}$ ergs.  In the present calculations, we generally
use somewhat different parameters, as discussed below.
%IVAN2: Is the above comment correct?  It seems inconsistent with some other comments.
%DAVID2:  Those values are estimated based on his experimental setup.  We used different 
%parameter values because the wave function approximation did not seem to work for driving 
%currents near the critical current.  That's why we used an I/I_c of around 0.85 and thus we
%used some different parameters than the martinis experiment.

To solve the time-independent Schr\"{o}dinger equation
\begin{equation}\label{eq:schrod1} \tilde{H}\Psi_i =
\tilde{E_i}\Psi_i
\end{equation}
for a given $I/I_c$, we first find the local minimum $\phi_0 =
\sin^{-1}(I/I_c)$ and corresponding potential $\tilde{U}_{min}
\equiv \tilde{U}(\phi_0)$ of the normalized washboard potential
$\tilde{U}(\phi) = \cos\phi - \frac{I}{I_c}\phi$.
We also find the closest local maximum $\phi_{max}$ and the
corresponding relative potential maximum, $\tilde{U}_{max} \equiv
\tilde{U}(\phi_{max})$.  For $I \sim I_c$, it is readily shown that
the barrier height
$\Delta\tilde{U} \equiv \tilde{U}_{max} - \tilde{U}_{min} \propto
\left[1 - I/I_c\right]^{3/2}$. 
The number of quasi-bound states $N_b(I)$ in the well near $\phi =
\phi_0$ is approximately $N_b(I) \sim \Delta
U(I)/[\hbar\omega_c(I)]$. The lifetime of a given quasi-bound state
in the well is also determined by this ratio.

To find the quasi-bound states, we expand the
solutions of eq.\ (\ref{eq:schrod1}) in the set of harmonic
oscillator states corresponding to the local minimum of $U(\phi)$
for the given $I/I_c$~\cite{cleland}.  These take the form
\begin{eqnarray}\label{psiharm}
\psi_n(\phi) = 2^{-n/2} (n!)^{-1/2} \bigg(\frac{1}{(\Delta\phi)^2
  \pi}\bigg)^{1/4}\times \\ \nonumber
 \times \exp\bigg(-\frac{(\phi-\phi_0)^2}{2(\Delta\phi)^2}\bigg)
  H_n\bigg(\frac{\phi-\phi_0}{\Delta\phi}\bigg),
\end{eqnarray}
where $H_n(\xi)$ is a Hermite polynomial, and
\begin{equation}
(\Delta\phi)^2 = [2e^2/\hbar C][\hbar C/(2eI_c)]^{1/2}\left(1 -
\frac{I^2}{I_c^2}\right)^{-1/4}.
\end{equation}

\noindent In this form, the orthonormal states are harmonic
oscillator eigenstates of the Hamiltonian
\begin{eqnarray}
{H}_{harm}(I) = U_{min} +\frac{\hbar I_c}{2e}\left(1 -
\frac{I^2}{I_c^2}\right)^{1/2}\times \\ \nonumber \times
\frac{(\phi-\phi_0)^2}{2} - E_C\frac{\partial^2}{\partial\phi^2},
\end{eqnarray}
where we have used the relation (\ref{eq:omegac}).

The solution $\Psi_i(\phi)$ of the time-independent
Schr\"{o}dinger equation is then expressed as $\Psi_i(\phi) = \sum_n a_n
\psi_n(\phi)$. In matrix form, the Schr\"{o}dinger equation becomes
\begin{equation}
\sum_n(\tilde{H}_{mn})a_n = \tilde{E}_ia_m,
\end{equation}
where $\tilde{H}_{mn} =
\int_{-\infty}^{\infty}\psi_m(\phi)\tilde{H}\psi_n(\phi)d\phi$.  In
our calculations, we include as many as the first $23$ harmonic
oscillator states in our matrix.    As a check of our procedure,
we have calculated
the lowest three energy eigenvalues $E_0$, $E_1$, and $E_2$, and
corresponding wave functions, for $I/I_c = 0.99$, approximately the
dc current studied in Ref.\ \cite{martinis}, using their
quoted values of $E_C$ and $E_J$, and from these compute
$\Delta E_{10}$, $\Delta E_{21}$ and the ratio $\Delta E_{21}/\Delta E_{10}$.  
We find $\Delta E_{10} = 3.86\times 10^{-17}$
erg, quite close to the value of $4.51\times 10^{-17}$ erg measured
in Ref.\  \cite{martinis}.   The difference may arise because the
actual values of $I/I_c$ and $C$ in the junction studied
experimentally differ slightly from the quoted experimental values.
We also calculate $\Delta E_{21}/\Delta E_{10} = 0.928$, quite close to
the value of 0.9 quoted for this quantity in Ref.~\cite{martinis}.
%IVAN2: are these numbers and remarks correct?
%DAVID2:  Yes.

Once we have these eigenstates, we determine how many of these are
"quasi-bound" by finding out how many satisfy $\tilde{E}_n <
\tilde{U}_{max}$.  In general, we have considered only values of
$I/I_c$ such that there are at least three quasi-bound states.  For
the parameters of Ref.\ \cite{martinis}, this condition is satisfied
up to $I/I_c \approx 0.99$.

\subsection{Current-Biased Josephson Junction with AC Voltage}

In the presence of an ac voltage $V_{ac}\sin\omega t$, the proper 
gauge-invariant phase difference takes the form\cite{tinkham}:
$\phi \rightarrow \phi +
\frac{2eV_{ac}}{\hbar\omega}\cos{\omega t}$.
The correspondingly modified Hamiltonian is
$H = -E_c\frac{\partial^2}{\partial \phi^2} - \frac{\hbar
I}{2\,e}\phi - \frac{\hbar I_c}{2\,e}\cos\left[\phi +
\frac{2eV_{ac}}{\hbar\omega}\cos\omega t\right]
- \frac{V_{ac}I}{\omega}\cos\omega t$.  This formulation makes clear that the
effects of the ac voltage are characterized by the frequency $\omega$ and by the dimensionless
variable $z = 2eV_{ac}/(\hbar\omega)$.
In the dimensionless form of Section IIA, the Hamiltonian
takes the form
\begin{eqnarray}\label{dimhamilt}
\tilde{H} = -\frac{E_C}{E_J}\frac{\partial^2}{\partial \phi^2} -
\frac{I}{I_c}\phi - \cos\left[\phi + \frac{\tilde{V}_{ac}}{\tilde{\omega}}\cos(\tilde{\omega}\tau)\right]
 \nonumber \\ - \frac{\tilde{V}_{ac}}{\tilde{\omega}}\frac{I}{I_c}\cos(\tilde{\omega}\tau),
\end{eqnarray}
where we have introduced the dimensionless frequency
$\tilde{\omega} = \frac{2e\omega}{I_c}$,
dimensionless time $\tau = \frac{I_c t}{2e}$,
and dimensionless ac voltage amplitude
$\tilde{V}_{ac} = \frac{V_{ac}}{R_0I_c}$,
where $R_0 = \hbar/(4e^2)$.

We are interested in terms in this Hamiltonian which will induce
transitions between the quasi-bound states discussed in Section IIA.
Since the last term in eq.\ (\ref{dimhamilt}) is a c-number, it will
not induce such transitions. The only relevant term is the Josephson
energy
$\cos\left[\phi +
\frac{\tilde{V}_{ac}}{\tilde{\omega}}\cos(\tilde{\omega}\tau)\right]$.
\noindent To extract the multiphoton transitions, we may express
this energy in terms of Bessel functions using the
expansions~\cite{abram} 
$\cos(z\cos(\theta))= J_0(z)
+ 2 \sum_{k=1}^\infty (-)^k J_{2k}(z)\cos(2k\theta)$ and
$\sin(z \cos(\theta)) = 2 \sum_{k=0}^\infty(-)^k J_{2k+1}(z)
\cos{(2k+1)\theta}$.
Writing $z = \tilde{V}_{ac}/\tilde{\omega}$ and $\theta =
\tilde{\omega}\tau$, and using a trigonometric identity, we can
easily show that
\begin{eqnarray}\label{expansion}
\cos(\phi + \frac{\tilde{V}_{ac}}{\tilde{\omega}}\cos(\tilde{\omega}\tau)) = J_0\bigg(\frac{\tilde{V}_{ac}}{\tilde{\omega}}\bigg)\cos\phi  \nonumber  \\
 + 2 \sum_{n=1}^\infty J_n\bigg(\frac{\tilde{V}_{ac}}{\tilde{\omega}}\bigg)\cos(\phi+n\pi/2)\cos(n\tilde{\omega}\tau).
\end{eqnarray}
Thus, in the presence of an ac voltage, the potential energy term in
the Schr\"{o}dinger equation is modified in two ways: (i) the
strength of the dc part is reduced by a factor of
$J_0(\tilde{V}_{ac}/\tilde{\omega})$, and (ii) there are an infinite
series of additional ac terms at frequencies $\tilde{\omega}$ and
all its harmonics.  These harmonic terms will induce the multiphoton
transitions.  Given these approximations, the full Hamiltonian takes
the form $H = H_0 + H_1$, where
\begin{equation}
H_0 = -E_c\frac{\partial^2}{\partial\phi^2} - \frac{\hbar I}{2e}\phi
- \frac{\hbar
I_c}{2e}J_0\left(\frac{2eV_{ac}}{\hbar\omega}\right)\cos\phi
\label{eq:h0}
\end{equation}
and
\begin{equation}
H_1 = 2\frac{\hbar I_c}{2e}\sum_{n=1}^\infty
J_n\left(\frac{2eV_{ac}}{\hbar\omega}\right)\cos(\phi+n\pi/2)\cos(n\omega
t). \label{eq:h1}
\end{equation}

The condition for the occurrence of an $n$-photon transition between
the ground and first excited state of the current-driven junction is
\begin{equation}
\Delta E_{10}(I, V_{ac}, \omega) = n\hbar\omega. \label{eq:nphoton}
\end{equation}
Here $\Delta E_{10}(I, V_{ac}, \omega)$ is the energy splitting between the
ground and first excited state when the ac voltage amplitude and
frequency are $V_{ac}$ and $\omega$.  We solve eq.\
(\ref{eq:nphoton}) by a self-consistent procedure.  For a given
$V_{ac}$ and $\omega$, the junction levels are obtained as solutions
of the time-independent Schr\"{o}dinger equation
$H(V_{ac},\omega)\Psi = E\Psi$, where $H(V_{ac},\omega)$ is given by
eq.\ (2) but with $E_J$ replaced by $E_JJ_0[2eV_{ac}/(\hbar\omega)]$
To find the strength of the $n$-photon transitions, we make an
initial guess for $\omega$, solve the dc Schr\"{o}dinger equation
using the potential $-E_J[J_0(2eV_{ac}/(\hbar\omega)]\cos\phi$, and
iterate until eq.\ (\ref{eq:nphoton}) is satisfied.
%IVAN, is this what you did?
%DAVID, Yes.
As in the previous section, we expand the quasi-bound eigenstates as
a linear combination of harmonic oscillator wave functions.  Thus,
we make an initial guess for $\omega$, calculate the eigenvalues,
and then change $\omega$ such that $n\hbar\omega = \Delta E_{10}(I, V_{ac}, \omega)$. 
The eigenvalues are recalculated using this new
$\omega$, and the procedure is repeated until the energy levels
remain unchanged to within an absolute error of $1.0\times
10^{-4}E_J$.
%IVAN: you mean that each eigenvalue remains unchanged to within plus or
%minus 0.0001 E_J?  I assumed this in writing the above sentence.
%DAVID, Yes, that is correct.
Although eq.\ (\ref{expansion}) is valid, in principle, for any
value of $z$, this procedure is not reasonable unless the strength
of the potential $[\hbar I_c/(2e)]J_0[2eV_{ac}/(\hbar\omega)] > 0$.
In practice, we find that, for our choice of parameters (see below), our
procedure leads to two or more bound states in the well only if
$\frac{2\,eV_{ac}}{\hbar \omega} < 0.58$.
%IVAN, what are E_C and E_J for your calculations which produce this limit?
%DAVID, The values that I used are E_C = 8.53333e-18 ergs
% and E_J  = 6.91121e-14 ergs

Given the energy levels, the n-photon transition rate
can be calculated, in principle, from the Fermi Golden Rule\cite{merzbach}:
\begin{eqnarray}
\tilde{\Gamma}_n(I, V_{ac}, \omega) & = &
 \frac{2\pi}{\hbar}|\langle 0 |V_n|1 \rangle|^2 \delta(E_s -
n\hbar\omega) \nonumber \\ & \equiv & \Gamma_n(I, V_{ac}, \omega)\delta(\Delta E_{10} -
n\hbar\omega),
\label{eq:golden}
\end{eqnarray}
where the perturbing potential for n-photon transitions is $V_n =
2J_n(\frac{2\,e\,V_{ac}}{\hbar \omega})\cos(\phi+n\pi/2)$.  The states $\langle0|$ and $|1\rangle$ are the ground and first
excited states of the current-biased junction in the presence of microwave radiation.

%REVISE: The remainder of this section discusses a way of including the dissipation due to 
%tunneling out of the higher quasi-bound state.

Eq.\ (\ref{eq:golden}) is suitable for describing transitions between bound states with
sharply defined energies. In reality, as already noted, the states of interest are only quasi-bound, since, for both the initial and the final state, the phase ``particle'' can tunnel out through the barrier of the washboard potential.  One way to take account of this tunneling is to describe both the initial and the final states by complex
energies $E_0 + i\gamma_0$ and $E_1 + i\gamma_1$.  The quantities $\gamma_0/\hbar$ and $\gamma_1/\hbar$ then represent the tunneling rates out of the states $|0\rangle$ and $|1\rangle$.
If this description is used, the n-photon transition rate is described by the second form
of the Fermi Golden Rule\cite{merzbach}, which represents the transition rate from an initial state of energy $E_0$ into a continuum
of final states:
\begin{equation}
\tilde{\Gamma}_n(I, V_{ac}, \omega) = \frac{2\pi}{\hbar}|\langle 0|V_n|1\rangle|^2\rho(\Delta E_{10}
- n\hbar\omega).
\label{eq:golden1}
\end{equation}
Here $\rho(E)$ is the density of final states.  If the final state is approximated by a complex energy
$E_1 + i\gamma_1$, then the quantity which replaces the delta function in eq.\ (\ref{eq:golden})
is
\begin{equation}
\rho(\Delta E_{10}-n\hbar\omega) = \frac{1}{\pi}\frac{\gamma_1}{(\Delta E_{10}-n\hbar\omega)^2 + \gamma_1^2},
\label{eq:golden1a}
\end{equation}
where the normalization is chosen so that $\int_{-\infty}^\infty\rho(E)dE = 1$.  The corresponding rate of {\it energy} absorption is
then 
\begin{equation}
\alpha = \sum_n n\hbar\omega\tilde{\Gamma}_n(I,V_{ac},\omega)(n\hbar\omega),
\label{eq:alphgold}
\end{equation}
where the factor of $n\hbar\omega$ denotes the fact that the energy absorbed
in an n-photon transition is $n\hbar\omega$. 

In order to use this formulation, we need to determine the width $\gamma_1$.  For the regime
studied in typical experiments, it should be sufficient to treat
$\gamma_1$ within the WKB approximation.  This approximation gives
\begin{equation}
\gamma_1 \sim \hbar\omega_0\exp\left(-2S/\hbar\right),
\label{eq:gam1}
\end{equation}
where $\omega_0$ is a suitable attempt frequency for escaping the well, and S is the
WKB action.  In this case, $S(E) = \int_{\phi_i}^{\phi_f}p_\phi d\phi$, where 
$p_\phi = -i\hbar(\partial/\partial\phi)$ is the momentum canonically conjugate to $\phi$, and
$\phi_i$ and $\phi_f$ are the values of $\phi$ at the left and right hand edges of the tunneling
barrier.  From the Hamiltonian (1) of the current-driven Josephson junction, we replace $p_\phi$ in
the WKB approximation by $\hbar\sqrt{[E - U(\phi)]/E_C}$.  Thus, we finally obtain (setting $E = E_1$
for the final state)
\begin{equation}
\frac{S(E)}{\hbar} \sim \int_{\phi_i}^{\phi_f}\sqrt{\frac{E - U(\phi)}{E_C}}d\phi,
\label{eq:action}
\end{equation}
%COMMENT: previously I thought this was 2S/\hbar - I believe the right hand side is just S/\hbar.
where $\phi_i$ and $\phi_f$ are determined by $U(\phi_i) = U(\phi_f) = E_1$.  We give some numerical estimates of $\gamma_1$, and of the corresponding multiphoton absorption, at the end of Section IV.  

Besides the escape rate from the metastable potential well, another source of line broadening is dissipation 
due to the finite shunt resistance $R$ of the underdamped Josephson junction.  We expect that this dissipation
will give rise to another contribution to $\gamma_1$, thereby further increasing the linewidths of the multiphoton
absorption lines.  The magnitude of this contribution obviously depends on the magnitude of $R$.

%REVISE: This is the end of my revisions for the last part of Section II.

\section{Inclusion of Dissipation using Bloch Equations}

%REVISE: the next paragraph has been changed somewhat to insure continuity with the changes above

The form of the result (\ref{eq:alphgold}) for multiphoton absorption in the presence of dissipative processes can also
be obtained using the analog of the Bloch equations familiar in NMR.  In this section, we describe this
method.  Related approaches have been discussed,
e.\ g., by Shevchenko {\it et al}\cite{shevchenko}.  Our approach goes beyond their
work because we directly compute the necessary matrix elements and thus obtain
explicit expressions for the absorption coefficient.

%REVISE: end of changed paragraph

We again consider the Hamiltonian $H = H_0 + H_1$ of eqs.\ (\ref{eq:h0}) and (\ref{eq:h1}), and
consider transitions only between the lowest two states of the Hamiltonian
(\ref{eq:h0}).   Since we are considering only two states, we may write the Hamiltonian 
(\ref{eq:h0}) in operator form as
$H_0 = -\frac{\Delta E_{10}}{2}\sigma_z + \mathrm{const}$, 
where $\Delta E_{10}(I, V_{ac},\omega)$ denotes the splitting between
the lowest two energy levels in the presence of an ac voltage of amplitude
$V_{ac}$ and frequency $\omega$.  The last term is a constant which can be chosen to vanish
by proper selection of the energy zero.
we choose our zero of energy so that this constant vanishes.  
$\sigma_i$  ($i = x$, $y$, $z$) are the three standard Pauli matrices.  
If we retain only that part of $H_1$ which produces transitions between $|0\rangle$ and
$|1\rangle$, then the time-dependent perturbation
(\ref{eq:h1}) can be written 
\begin{equation}
H_1 = \sum_{n=1}^\infty 2 J_n\left(\frac{2eV_{ac}}{\hbar\omega}\right)
\langle 0 |\cos(\phi + n\pi/2)|1\rangle\cos(n\omega
t)\sigma_x. \label{eq:hh1}
\end{equation}
In writing the Hamiltonian in this form, we are not only considering
just two levels but also are using the fact that the
diagonal matrix elements of the form
$\langle 0|H_1(t)|0\rangle$ and $\langle 1|H_1(t)|1\rangle$ vanish.
%IVAN2: I am not sure if this is really true.  Perhaps we should
%say that we neglect these diagonal matrix elements.  What do you
%think?  Is this statement true?
%DAVID2:  In the calculation, those off diagonal elements do vanish.  So
%DAVIS2:  yes, it is true.

In the absence of relaxation processes, the Heisenberg equations of
motion can be used to obtain equations of motion for expectation
values of any operators.  We write the time-dependent wave function as
$|\psi(t)\rangle = a(t)|0\rangle + b(t)|1\rangle$, where
$|a|^2 + |b|^2 = 1$.   We also write the Hamiltonian as
$H = H_0 + H_1 = -{\bf B}\cdot {\bf \sigma}/2$,
where ${\bf \sigma}$ is the ordered triple of Pauli matrices, and
${\bf B}$ is an effective magnetic field, which is the sum of a time-independent 
$B_z{\bf \hat{z}}$ and a time-dependent part $B_x(t){\bf \hat{x}}$, with 
\begin{equation}
B_z = \Delta E_{10},
\label{eq:bz}
\end{equation}
\begin{equation}
B_x(t) =  \frac{\hbar I_c}{2e}\sum_{n=1}^\infty 4 J_n\left(\frac{2eV_{ac}}{\hbar\omega}\right)
\langle 0 |\cos(\phi + n\pi/2)|1\rangle\cos(n\omega t),
\label{eq:bx}
\end{equation}
and $B_y = 0$.

The Heisenberg equations of motion for the expectation value of any 
operator ${\cal O}$ are
$i\hbar\frac{d}{dt}\langle {\cal O} \rangle = i\hbar
\frac{d}{dt}\langle \psi|{\cal O}|\psi\rangle = \langle \psi|[{\cal
O}, H]|\psi\rangle$. 
For the present Hamiltonian $H = H_0 + H_1$, including only the two lowest energy levels, they reduce to the well-known form
\begin{equation}
\frac{d}{dt}{\bf M} = \frac{1}{\hbar}{\bf M} \times {\bf B},
\label{eq:bloch0}
\end{equation}
where ${\bf M} = \langle {\bf \sigma}\rangle$, the brackets denoting a quantum-mechanical 
expectation value.

The form of eq.\ (\ref{eq:bloch0}) suggests that, to include dissipative processes, we should
generalize this equation so that it has the same form as the Bloch equations of magnetic resonance:
\begin{eqnarray}
\frac{d}{dt}{\bf M}_\perp & = & \frac{1}{\hbar}\left({\bf M}
\times {\bf B}\right)_\perp -
\frac{{\bf M}_\perp}{\tau_\phi} \nonumber \\
\frac{d}{dt}M_z & = & \frac{1}{\hbar}\left({\bf M} \times {\bf
B}\right)_z - \frac{(M_z - M_0)}{\tau_r}.
\label{eq:bloch1}
\end{eqnarray}
Here $\tau_\phi$ and $\tau_r$ are relaxation times analogous to
$T_1$ and $T_2$ in magnetic resonance theory, ${\bf
M}_\perp \equiv (M_x, M_y)$, and $M_0\hat{z}$ is the equilibrium
value of $\langle \sigma \rangle$ to which ${\bf M}$ reverts when the time-dependent
perturbation is turned off [see, e.\ g., Ref.\ \cite{slichter}).  

%NEED REFERENCES HERE.  

Given the solution for ${\bf M}(t)$, we can calculate the absorbed power
by adapting a standard approach used in NMR
calculations\cite{slichter}, namely, we solve
for $M_x(t)$ in the limit when $B_x(t)$ is weak. In this regime,
$M_x(t)$ can be obtained in closed form by superimposing the solutions 
arising from each separate frequency $\Omega$ of $B_x(t)$ in eq.\ (\ref{eq:bx}).  For
small $B_x$, $M_z \sim M_0$ through first order in $B_x$
Writing the three Bloch equations in
component form,
%\begin{eqnarray}
%\frac{dM_x}{dt} & = & \frac{1}{\hbar}M_yB_z - \frac{M_x}{\tau_\phi} \nonumber \\
%\frac{dM_y}{dt} & = & \frac{1}{\hbar}(M_zB_x - M_xB_z) - \frac{M_y}{\tau_\phi} \nonumber \\
%\frac{dM_z}{dt} & = & - \frac{1}{\hbar}M_yB_x - \frac{(M_z -
%M_0)}{\tau_r}.
%\end{eqnarray} 
and also assuming that both $M_x$ and $M_y$ vary with time as $\exp(-i\Omega t)$, 
we obtain
\begin{eqnarray}
-i\Omega M_x & = & \frac{1}{\hbar}M_yB_z - \frac{M_x}{\tau_\phi}; \nonumber \\
-i\Omega M_y & = & \frac{1}{\hbar}(M_0B_x - M_xB_z) -
\frac{M_y}{\tau_\phi}.
\end{eqnarray}
The resulting solution for $M_+ \equiv M_x + iM_y$ is readily found to be
\begin{equation}
M_+(t) = \frac{(B_x/\hbar)M_0e^{-i\Omega t}}{(B_z/\hbar) -
\Omega - i/\tau_\phi}.
\end{equation}

For the present problem, the transverse field is a sum of terms of
the form $B_{nx}\cos(n\omega t) \equiv 
(B_{nx}/2)(e^{-i\omega t} + e^{+i\omega t})$, which
induces a transverse magnetization
\begin{eqnarray}
M_{n,+}(t) & = &(B_{nx}/2\hbar)M_0\times \nonumber \\
& \times &\left(\frac{e^{-in\omega
t}}{B_{nz}/\hbar - n\omega - i/\tau_\phi} + \frac{e^{in\omega
t}}{B_z/\hbar + n\omega - i/\tau_\phi}\right). \nonumber \\
\end{eqnarray}
Hence $M_{nx}(t) \equiv \mathrm{Re}M_{n,+}(t)$ satisfies
\begin{eqnarray}
\frac{2\hbar M_0M_{n,x}}{B_{nx}}  =   \nonumber \\
 \frac{\cos(n\omega
t)(B_{nx}/\hbar - n\omega)+\sin(n\omega
t)(1/\tau_\phi)}
{(B_{z}/\hbar-n\omega)^2+1/\tau_\phi^2} \nonumber \\
  +  \frac{\cos(n\omega
t)(B_{nx}/\hbar+n\omega)-\sin(n\omega
t)(1/\tau_\phi)} {(B_{z}/\hbar+n\omega)^2
+1/\tau_\phi^2}.
\end{eqnarray}
Finally, the total absorption can be written
\begin{equation}
\alpha = \langle\frac{d}{dt}\langle H_1(t)\rangle\rangle_t
\end{equation}
where the outer brackets denote a time average.  
For our time-dependent
perturbation, this expression is readily shown to be equivalent to
\begin{eqnarray}
\alpha = \sum_{n=1}^\infty
M_0\left(\frac{B_{nx}}{\hbar}\right)^2\frac{n\omega}{4\tau_\phi}\times \nonumber \\
\times \left[\frac{1}{(B_z/\hbar-n\omega)^2 +1/\tau_\phi^2}
+\frac{1}{(B_z/\hbar+n\omega)^2 + 1/\tau_\phi^2}\right].
\label{eq:abs}
\end{eqnarray}
In terms of the
parameters of the two-level system, we have 
\begin{equation}
\alpha = \sum_{n = 1}^\infty \alpha_n,  \label{eq:abs1}
\end{equation}
where
\begin{equation}
\alpha_n =
M_0\frac{\hbar I_c^2}{2e^2}J_n\left(\frac{2eV_{ac}}{\hbar\omega}\right)^2\frac{n\omega}{4\tau_\phi}
|\langle 0|\cos(\phi + n\pi/2)|1\rangle|^2{\cal
L}(n\omega), \label{eq:abs2}
\end{equation}
where ${\cal L}(n\omega)$ is a Lorentzian given by
\begin{eqnarray}
{\cal L}(n\omega) = \frac{1}{(\Delta E_{10}/\hbar-n\omega)^2
+1/\tau_\phi^2} \nonumber \\ +\frac{1}{(\Delta E_{10}/\hbar+n\omega)^2 +
1/\tau_\phi^2}. \label{eq:abs3}
\end{eqnarray}
Thus, in this approximation, the total absorption is just
a sum of Lorentzians.  The quantity $M_0$, undetermined in this calculation, is given
by $2p - 1$, where $p$ is the probability that the junction is to be found in the lower
of the two states, $|0\rangle$.   The {\em integrated strength} of the n-photon absorption
line is
\begin{equation}
\int_{-\infty}^\infty \alpha_n(\omega) = M_0 \frac{2\pi}{\hbar}\tau_\phi B_{nx}^2B_{nz} = \Gamma_n.
\label{eq:intalpha}
\end{equation} 
where $\Gamma_n$ is the coefficient of 
eq.\ (\ref{eq:golden}).  
%REVISE: next sentence added in revisions.
Eqs.\ (\ref{eq:abs1}), (\ref{eq:abs2}), and (\ref{eq:abs3}) are basically equivalent
to the result (\ref{eq:alphgold}) obtained in the previous section.  The quantity $\hbar/\tau_\phi$
is analogous to the quantity $\gamma_1$ in eq.\ (\ref{eq:golden1a}).

In principle, a more accurate solution could be obtained by solving
the Bloch equations directly for the full time-dependent Hamiltonian
arising from the microwave voltage.  This would lead to a more complicated absorption
lineshape, which would probably depend on $\tau_r$ as well as $\tau_\phi$.

\section{Numerical Results}

We now describe our calculated $n$-photon transition
rates $\Gamma_n(I, V_{ac}, \omega)$, in the absence of dissipation, 
for $n$ ranging from $1$ to $5$. In each calculation, we start by choosing
$I$, $V_{ac}$ and $n$.  The frequency and energy-level
splitting are then determined self-consistently, as described above,
and the transition rate $\Gamma_n$ is calculated [see eq.\ (\ref{eq:golden})].  In all calculations, we choose
$E_C = 8.53\times 10^{-18}$ erg, $E_J = 6.91 \times 10^{-14}$ erg,
and hence $E_C/E_J = 1.235 \times 10^{-4}$.  For this choice, there
are three quasi-bound states at $I/I_c \sim 0.85$, and the ratio
$\Delta E_{21}/\Delta E_{10} \sim 0.928$.
%IVAN2: this ratio is different from what we stated earlier in the paper (around 0.92).
%Which one is right?  How about the other numbers?  Can you check this?
%DAVID2: It should have been 0.928.  I had left out the 2.  Sorry.

%, similar to the experiment of Ref.\
%\cite{martinis}.
%IVAN, could you fill in the value of E_C/E_J which you actually used in
%these calculations?  Also, I more or less made up the last sentence; is it
%correct?
%DAVID, That value is not correct.  It is actually around 0.98, which I put in to the
%text and took out the reference part of the sentence.

\begin{figure}
\begin{center}
\includegraphics[angle = -90, width = 0.5\textwidth]{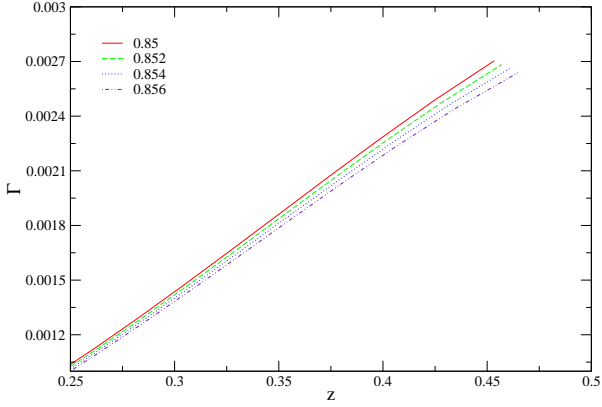}
\caption
{ \small (Color online.)  Single photon transition rate $\Gamma_1$ (in units of $E_J^2/\hbar$) as a function of $z =
\frac{2eV_{ac}}{\hbar \omega}$ for currents ranging from $I/I_c =
0.85$ to $0.856$.  The red (dark gray full) line, green (light gray dashed) line, blue (gray dotted) line, and violet (gray dash-dotted) line correspond to z = 0.85, 0.852, 0.854, and 0.856 respectively.
The same color and grayscale codes are used in Figs.\ 3-11.}\label{one_a}
\end{center}
\end{figure}

\begin{figure}
\begin{center}
\includegraphics[angle = -90, width =
0.5\textwidth]{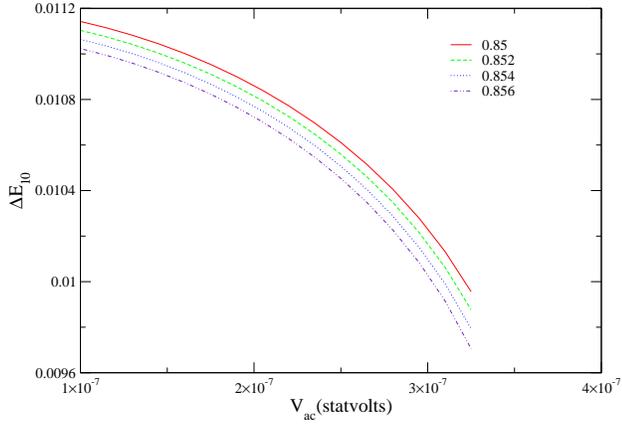} \caption { \small
(Color online.)  Energy level splitting $\Delta E_{10}$ (in units of $E_J$ )between the ground and first excited
  states as a function of $V_{ac}$ for currents $I/I_c$ ranging from $0.85$ to
  $0.856$ for the self-consistently determined frequency $\omega = \Delta E_{10}/\hbar$.  Parameters
  are the same as in Fig.~\ref{one_a}.}\label{one_c}
%IVAN1: I know that the journal will want the horizontal axes to be drawn with conventional
%exponential notation, i. e. 10^{-7} rather than e-07.  Maybe you could draw the horizontal axis
% here and in later figures as, say, 1     2         3            4 x 10^{-7} (that is, just put in
%the exponential once).  What do you think?   
%Also, I suggest changing the notation (here and in later figures) 
%from \Delta E_{10} to just E_{10} - this is what I did in
% the text.  But if it is too much of a nuisance, I can just change the notation in the text.
\end{center}
\end{figure}

In Fig.\ \ref{one_a}, we show $\Gamma_1$ as a function of 
$z = 2eV_{ac}/(\hbar \omega)$ for a narrow range of driving
currents, $0.85 < I/I_c < 0.856$. 
%IVAN, why did you use such a small range of driving currents?
%The same transition rates are plotted as functions of $V_{ac}$ in
%Fig.\ (\ref{one_b}).
%DAVID, I used such a small range to stay within the region where
% the depth of the well was non-zero.
The frequency is chosen so that $\hbar\omega = \Delta E_{10}(I, V_{ac}, \hbar\omega)$.  
For the chosen range of $I$, $\Gamma_1$ is quite insensitive to $I$.
In Fig.\ \ref{one_c}, we plot $\Delta E_{10}(I, V_{ac}, \omega)$
as a function of $V_{ac}$ at a frequency satisfying
$\hbar\omega = \Delta E_{10}$.
%IVAN, I added this last comment - is it correct?
%DAVID, Yes, that is correct.
The splitting decreases roughly quadratically with increasing
$V_{ac}$.  This behavior is expected from the Bessel-function dependence
of the potential strength, since $J_0(z)$ decreases approximately
quadratically with $z$, resulting in a shallower potential
well with increasing $z$ for fixed $I$.

\begin{figure}
\begin{center}
\includegraphics[angle = -90, width =
0.5\textwidth]{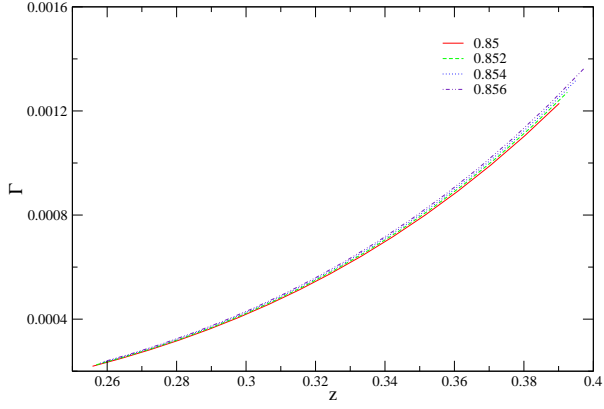} \caption { \small (Color online.)  Two
photon transition rate $\Gamma_2$ plotted as a function of $z =
\frac{2eV_{ac}}{\hbar \omega}$ for currents ranging from $I/I_c =
0.85$ to $0.856$ and the self-consistently determined frequency
$\omega = \Delta E_{10}/(2\hbar)$.}\label{two_a}
\end{center}
\end{figure}

\begin{figure}
\begin{center}
\includegraphics[angle = -90, width = 0.6\textwidth]{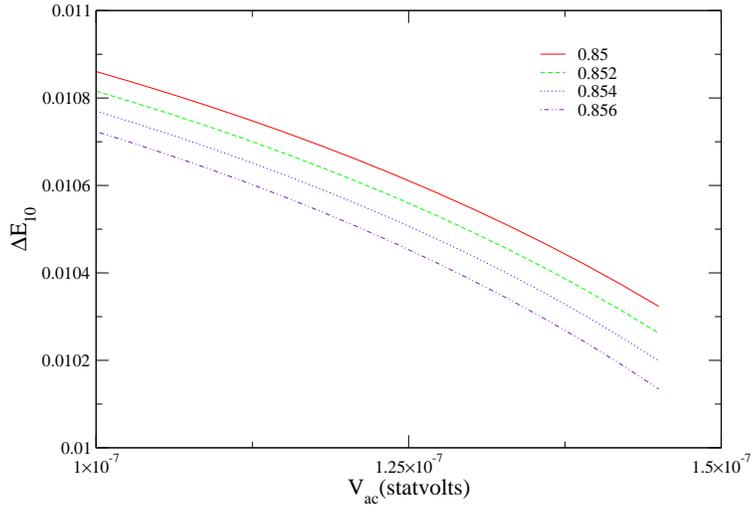}
\caption
{ \small (Color online.)  Energy level splitting $\Delta E_{10}$ as a function of $V_{ac}$ for currents ranging $I/I_c = 0.85$ to
  $0.856$ and a frequency such that $\omega = \Delta E_{10}/(2\hbar)$.}\label{two_c}
\end{center}
\end{figure}

In Figs.\ \ref{two_a} and \ref{two_c}, we show corresponding results
for two-photon transitions (n = 2), all other parameters being kept
the same as in Figs.\ \ref{one_a} and \ref{one_c}.
%IVAN - did you ever explain those parameters?  Specifically, what is the
% ratio of U to E_J - don't you need to specify this?
%DAVID, I think those are the values that we describe above for E_c and E_J
In these plots, we have considered smaller values of
$V_{ac}$ than for $n = 1$.   We do so because of the condition 
$n\hbar\omega = \Delta E_{10}(I, V_{ac}, \omega)$ which must be satisfied for
$n$-photon absorption.  Two-photon absorption thus occurs at roughly half the
frequency of one-photon absorption, as expected.  $\Delta E_{10}$ is more sensitive to $V_{ac}$
at such frequencies than it would be for one-photon absorption.  We must also
keep $V_{ac}$ well below the limit $V_{ac} = 0.58 n \hbar\omega/(2e)$, above which  
there are no quasi-bound states. 

%\begin{figure}
%\begin{center}
%includegraphics[angle = 0, width = 0.85\textwidth]{blocheqsol.ps}
%\caption {\small Multiphoton absorption coefficient
%$\alpha(\omega)$, as calculated from a solution of the Bloch
%equations in the weak-field approximation.  The parameters used are
%$E_{10} = 0.01\hbar I_c/(2e)$, $I/I_c = 0.85$, and
%$1/\tau_\phi = 0.1 E_{10}/\hbar$ }
%\end{center}
%\end{figure}

\begin{figure}
\begin{center}
\includegraphics[angle = -90, width = 0.5\textwidth]{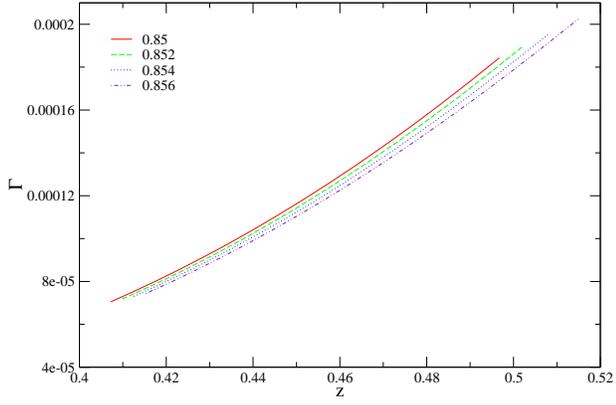}
\caption
{ \small (Color online.)  Same as Fig.~\ref{two_a} but for the three-photon transition rate $\Gamma_3$.}\label{three_a}
\end{center}
\end{figure}

\begin{figure}
\begin{center}
\includegraphics[angle = -90, width = 0.5\textwidth]{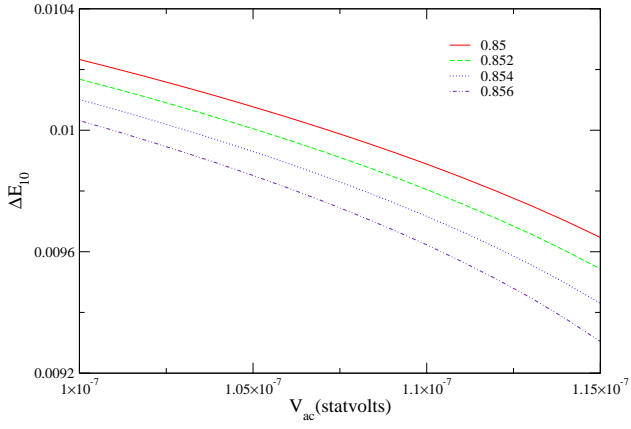}
\caption
{ \small (Color online.)  Same as Fig.~\ref{two_c} but for three-photon transitions.}\label{three_c}
\end{center}
\end{figure}

\begin{figure}
\begin{center}
\includegraphics[angle = -90, width = 0.5\textwidth]{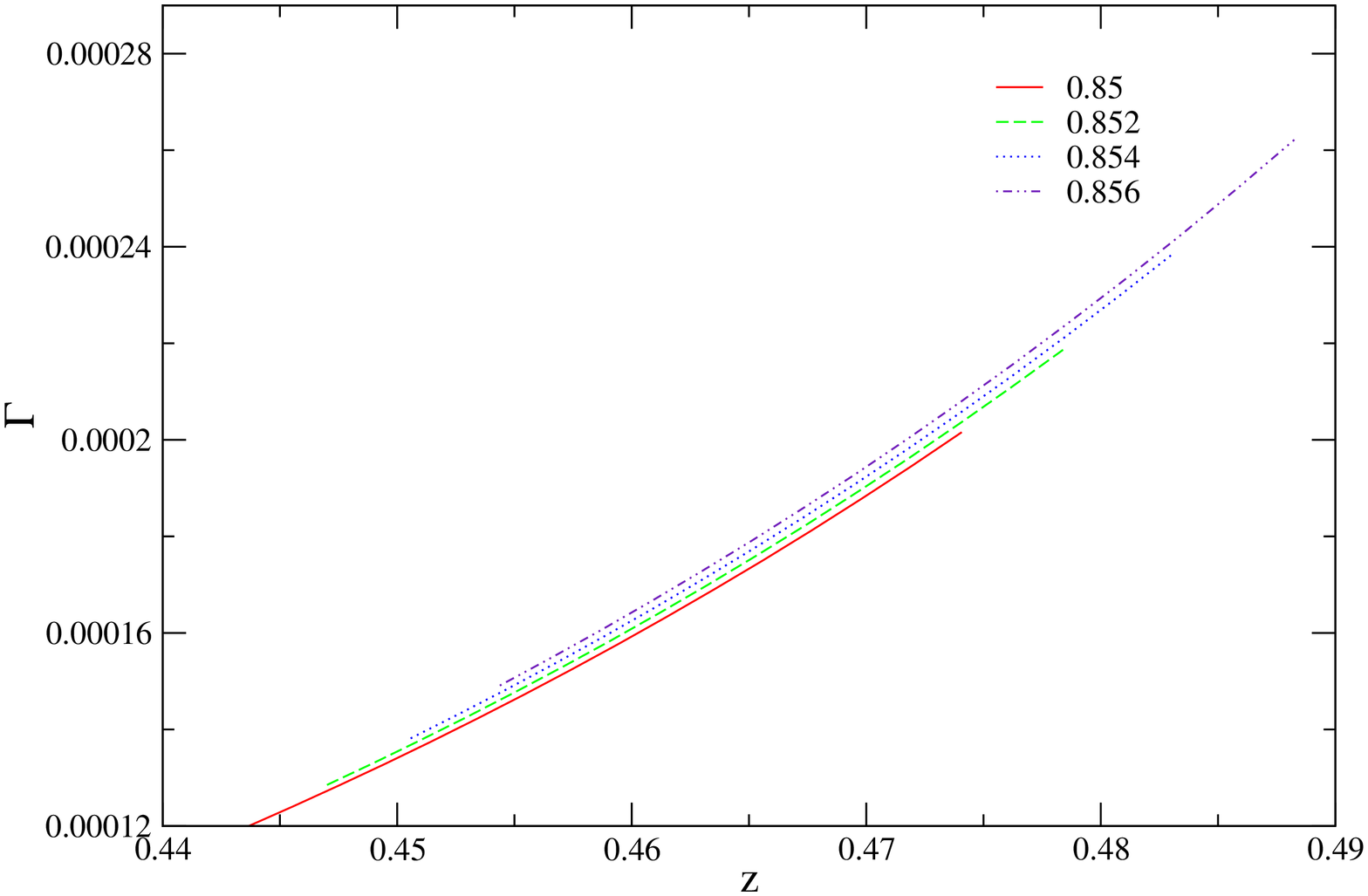}
\caption
{ \small (Color online.)  Same as Fig.~\ref{two_a} but for four-photon transitions}\label{four_a}
\end{center}
\end{figure}

\begin{figure}
\begin{center}
\includegraphics[angle = 0, width = 0.5\textwidth]{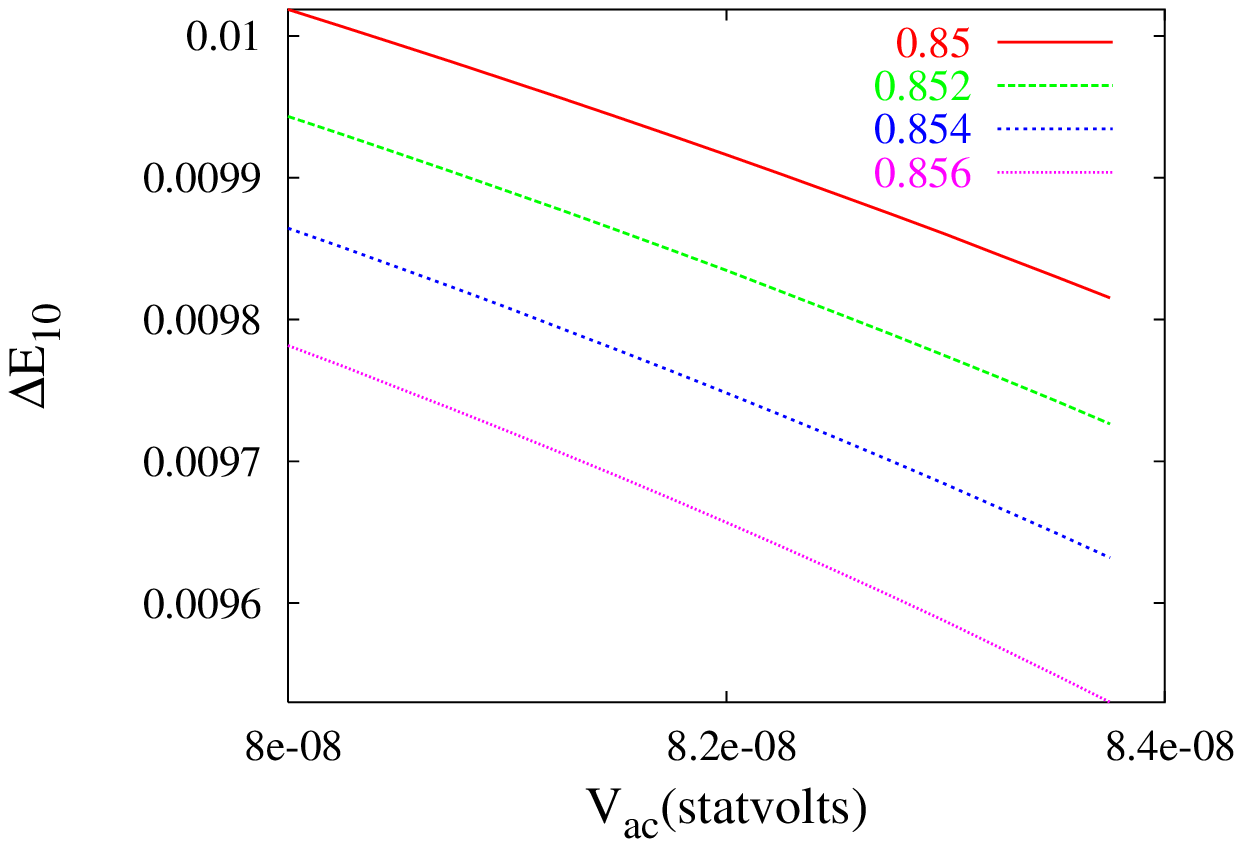}
\caption
{ \small (Color online.)  Same as Fig.~\ref{two_c} but for four-photon transitions.}\label{four_c}
\end{center}
\end{figure}

\begin{figure}
\begin{center}
\includegraphics[angle = -90, width = 0.5\textwidth]{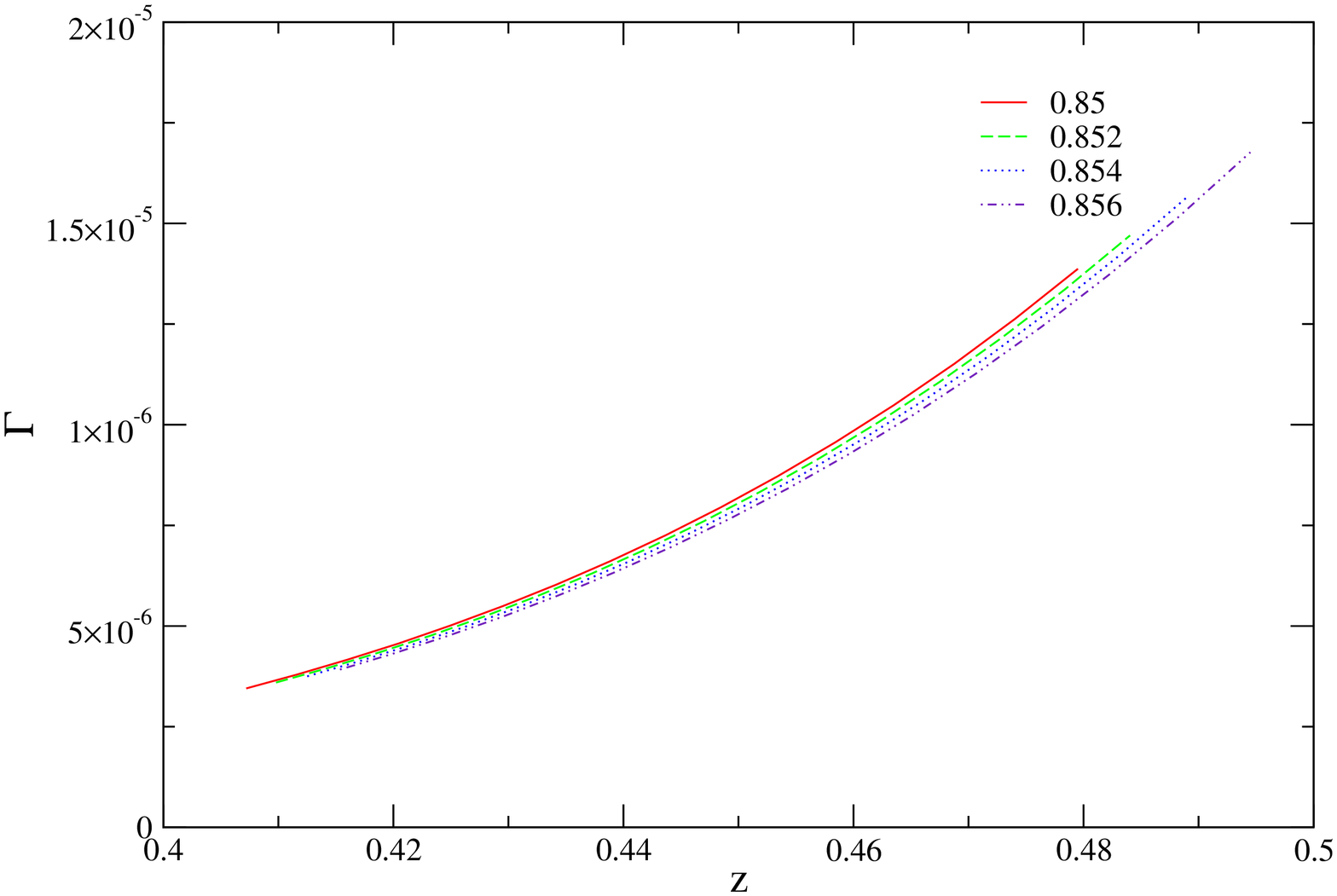}
\caption
{ \small (Color online.)  Same as Fig.~\ref{two_a} but for five-photon transitions. }\label{five_a}
\end{center}
\end{figure}

\begin{figure}
\begin{center}
\includegraphics[angle = -90, width = 0.5\textwidth]{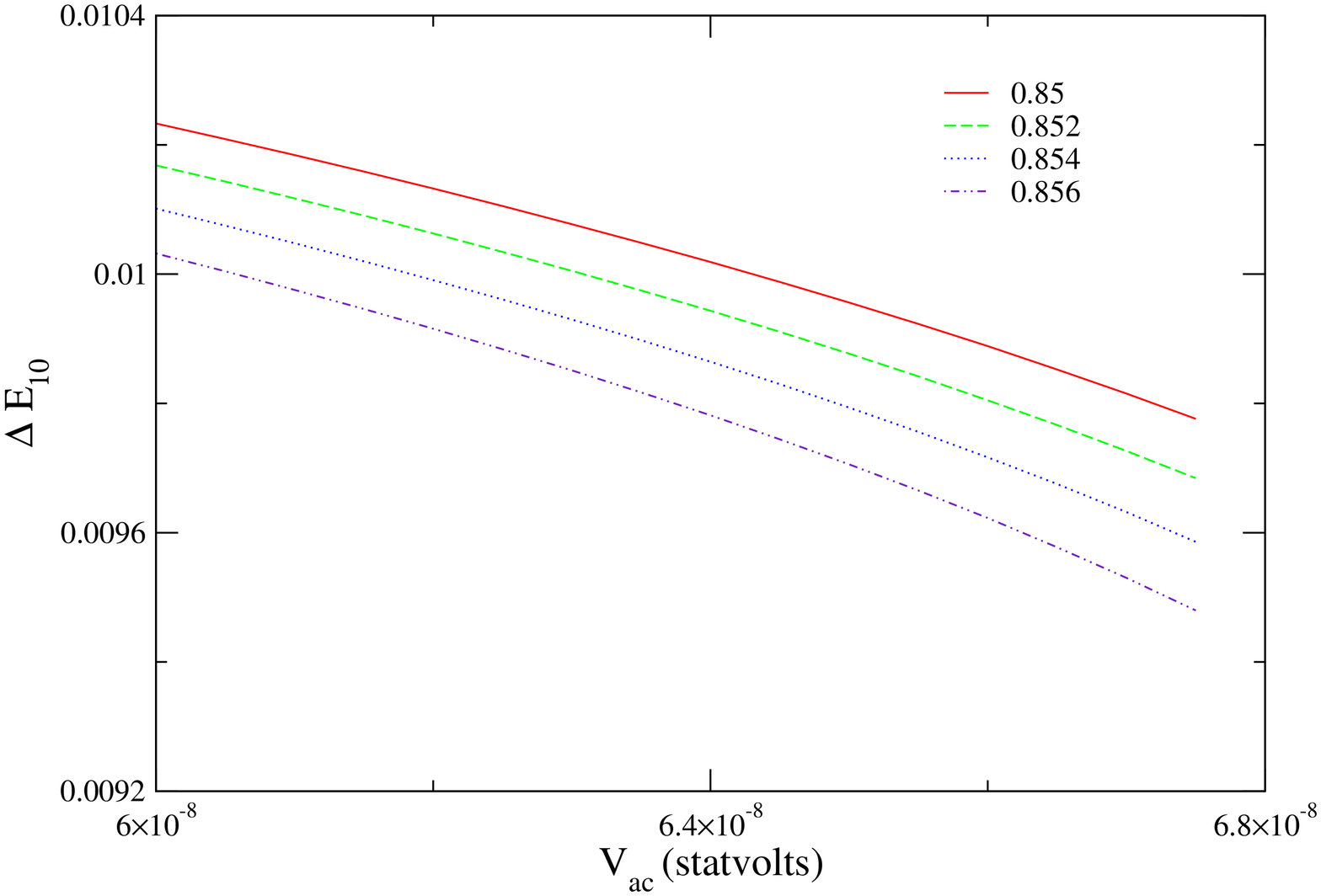}
\caption
{ \small (Color online.)  Same as Fig.~\ref{two_c} but for five-photon transitions.}\label{five_c}
\end{center}
\end{figure}

The corresponding results for three-photon transitions are shown in
Figs.~\ref{three_a} and~\ref{three_c}.  In general $\Gamma_3 < \Gamma_1$, as expected, since
the corresponding matrix element involves $J_3(z)$ instead of $J_1(z)$.   Furthermore, of course,
for a given $I$ and $V_{ac}$, the frequency required to produce the $n = 3$ transition
is smaller than that for $n = 1$.  We have also considered smaller values of $V_{ac}$ than for 
$n = 1$, because the sensitivity of $\Delta E_{10}$ to $V_{ac}$ is greater for $n = 3$ than for
$n = 1$.  In Figs.~\ref{four_a} -\ref{five_c} we show the corresponding results for $n = 4$ and
$n = 5$ transitions. 

All the above results make clear a noticeable difference
between even and odd transitions: for $n$ even, $\Gamma_n$ {\em increases}
with increasing $I$ at fixed $z$, whereas for $n$ odd, $\Gamma_n$
{\em decreases} with increasing $I$.  (We have not considered a very large range of
$I$ in calculating these rates.)
%IVAN2: I added the above sentence.  See if you think the rest of this paragraph makes sense.
%DAVID2:  Yes it makes sense.
This difference is easily
understood from the matrix elements of the time-dependent
perturbation which enter equation (\ref{eq:golden}) for the
transition rate.  Specifically, the argument of the cosine term has
a phase shift of $n\pi/2$, so that the perturbing potential varies
as $\sin\phi$ for odd $n$.   Because the perturbation is even in $\phi$ for even $n$, $\Gamma_n$ vanishes for even $n$ at $I = 0$, and hence should increase within increasing $I$, as we observe.  By
contrast, the perturbation is odd in $\phi$ for odd $n$, so that
$\Gamma_n$ for odd $n$ is finite at $I = 0$ but decreases with
increasing $I$.  Thus, we interpret this even-odd effect as
arising from the different parities of the time-dependent perturbation for
$n$ even and $n$ odd.  

Figs.\ 2 - 10 also show that, for fixed $z = 2eV_{ac}/(\hbar\omega)$ and fixed
$I/I_c$, $\Gamma_n$ generally decreases with increasing $n$.  Comparison of the
behavior for different $n$ is somewhat difficult, because the relevant value of the parameter
$z = 2eV_{ac}/(\hbar\omega)$ is different for each $n$.
%IVAN2: actually, it is a little difficult to compare different n's because
%the calculations are plotted as a function of z rather than V_{ac} directly
This behavior is consistent with that observed in the experiments of
Ref.\ \cite{martinis}.  These authors find that, when the first
excited state is populated by $n$-photon transitions, the apparent
lifetime increases with increasing $n$. This increased lifetime is
simply the consequence of a smaller transition rate for
the higher-$n$ transitions.

%Finally, in Fig.\ 12, we show the absorption spectrum in the
%presence of dissipation, as obtained using eqs.\
%(\ref{eq:abs1})-(\ref{eq:abs3}).  The calculations are carried out
%using the junction parameters described above, with $I/I_c = 0.85$
%and $V_{ac} = 2.5 \times 10^{-7}$ statvolts.  For this choice,
%$\Delta E_{10} \sim .01(\hbar I_c/(2e)$.
%IVAN, I assume these are the units?
%We also make the arbitrary assumption that $\hbar/\tau_\phi =
%0.1 E_{10}$.  For this choice, we can see a number of
%multiphoton absorption lines in Fig.\ 12.
%IVAN1: I suggest having one more figure showing what happens when one puts in
%Lorentzian damping, as discussed in the previous section.  What do you
%think?
%IVAN2: I don't think this figure is necessary, so I removed the discussion
%about it.  Let me know if you think it should go back in.
%DAVID2:  I think it is fine to leave it out.

%REVISE: In the next two paragraphs, I describe a model calculation including dissipation, using
%the broadened levels and the WKB approximation as discussed at the end of Section II.

Finally, we estimate how these n-photon transition rates are affected when we include 
reasonable estimates of level broadening due to tunneling of the phase ``particle'' through
the barrier.  We use the formalism of eqs.\ (\ref{eq:golden1}) - (\ref{eq:action}).  Using the values of $E_C$ and $E_J$ given above (which lead to $E_C/E_J = 1.235 \times 10^{-4}$), we obtain estimates for $\exp[-2S(E_1)/\hbar]$ given in Table 1 for $I/I_c = 0.85$, $0.90$, and $0.95$.  The corresponding width $\gamma_1$ of the Lorentzian line shape is given, as a fraction of $\hbar\omega_c(I)$, in the last column.
In calculating $S(E_1)$, we have estimated $E_1$, the energy of the second-lowest quasibound state,
by expanding the potential in the Hamiltonian (1) through third order in $\phi -\phi_0$, and have included the
cubic term as a perturbation to the resulting harmonic Hamiltonian, through second order in perturbation
theory.  The resulting values of $E_1$ are also shown in the Table.

In order to illustrate a typical multiphoton absorption spectrum in the presence of this
dissipation, we have carried out a simplified model calculation of the total absorption spectrum,
$\tilde{\Gamma}_{tot}(I, V_{ac}, \omega) = \sum_{n = 1}^{n_{max}}\tilde{\Gamma}_n(I, V_{ac}, \omega)$,
where $\tilde{\Gamma}_n$ is given by eqs.\ (\ref{eq:golden1}) and (\ref{eq:golden1a}).  The results are shown in Figs.\ 12 and Fig.\ 13 for
$I/I_c = 0.95$, $E_C = 8.53 \times 10^{-18}$ erg, and $E_J = 6.91 \times 10^{-14}$ erg as earlier in
this paper.  We assumed $I/I_c= 0.95$ and considered two ac amplitudes given by 
$2eV_{ac}/\hbar\omega_0(I) = 0.2$ and $0.4$.  We have included all multiphoton terms through 
$n_{max} = 7$.  For this calculation,
we have calculated both the energies $E_0$ and $E_1$ and the matrix elements  
$\langle 0| V_n|1\rangle$ by including only the lowest order anharmonic corrections to $E_0$, $E_1$, $|0\rangle$
and $|1\rangle$, the perturbation being the terms in the potential of third order in $\phi - \phi_0$.  In the interests of simplifying this calculation, we have also neglected the dependence of the Josephson well depth on the ratio $z = 2eV_{ac}/(\hbar\omega)$ [eq.\ (11)].  In practice,
this means that this model calculation becomes increasingly inaccurate at low frequencies.  For our choice
of $V_{ac}$, this dependence on $V_{ac}$ should probably be included for $\hbar\omega < 0.2 \Delta E_{10}$.  With these assumptions, however, both the energies and the matrix elements can be computed in closed form in terms of $E_J$, $E_C$ and $I/I_c$.  The results show that the strength of the lines falls off rapidly with increasing $n$ for our chosen value of $V_{ac}$.

In these calculations, and with our estimate of $\gamma_1$, we have not 
included the effects of the shunt resistance $R$, which is always present
in a realistic Josephson junction.  This shunt resistance will further broaden
the multiphoton absorption spectrum, but is distinct from the tunneling through the barrier treated above.
For the junctions studied in Ref.\ \cite{wallraff2}, $1/[\omega_p(0)\tau] < 5 \times 10^{-4}$, using their parameters
of critical current density $j = 1.1$ kA/cm$^2$, junction area $\sim (5.5 \mu)^2$, $C \sim 1.6$ pF, and
shunt resistance $R > 500 \Omega$.  Here $\tau = RC$ is the relaxation time associated with resistive dissipation
within the junction.  If we assume the same value of $1/[\omega_p(0)\tau]$ for our model junctions, the broadening
of the multiphoton lines at $I/I_c = 0.95$ would be about ten times smaller due to this source of dissipation
than that due to tunneling through the barrier.  For other junctions, of course, this might be the dominant source
of broadening.

\section{Discussion}

We have described a simple model for multiphoton transitions in a
current-biased Josephson junction, in the presence of microwave
irradiation.  The model allows calculation of the transition rates
between the ground and excited states as a function of dc bias
current, ac voltage amplitude, and frequency.   In the absence of
damping, the absorption occurs as a series of delta-function lines
satisfying the energy-conservation requirement $n\hbar\omega = 
\Delta E_{10}(I, V_{ac}, \omega)$.  The lines are broadened by
dissipative processes.  We calculate this broadening using the analog
of the Bloch equations as applied to the lowest two levels of the
junction.   This directly gives the absorption coefficient $\alpha(\omega)$,
which is approximately a sum of Lorentzian absorption lines
centered around frequencies corresponding to n-photon absorption.

Experimentally, multiphoton transitions are detected via enhanced
tunneling from the excited quasi-bound states through the barrier of
the washboard potential into the continuum\cite{martinis}.   In the
present work, we calculate the steady-state absorption rate $\alpha(\omega)$, using the Bloch equations.  By conservation of energy, we expect that this
rate should equal the rate of tunneling through the barrier from the state $E_1$.  Since our calculated $\alpha_n(\omega)$ generally decreases with increasing $n$, this would imply that the rate of tunneling through the barrier would also decrease with increasing $n$, as is reported in experiments\cite{martinis}.  However, we have not attempted to calculate the relevant phase relaxation time $\tau_\phi$ which determines the width of the Lorentzian peaks in this approximation.
%IVAN2: I slightly reworded the above paragraph.
%DAVID2: I think this is fine.

The calculation of $\alpha(\omega)$ via the Bloch equations is presumably
derivable from the master-equation approach\cite{larkin} previously used to discuss the time-dependence of level occupation numbers in Josephson junctions.  In the present work, we also explicitly calculate the matrix
elements needed to calculate $\alpha_n(\omega)$, which allows us to 
explicitly estimate the multiphoton absorption rate as a function 
of $I$, $V_{ac}$, and $\omega$. 
%IVAN2: I also reworded this paragraph.  What do you think?  Shall we
%keep this paragraph?
%DAVID2: Yes, keep it.  

Our calculations have several characteristic qualitative features.  For example, in agreement with experiment\cite{martinis}, the $n$-photon absorption rate, for a given $I$ and $V_{ac}$, generally decreases with
increasing $n$.   Also, if the damping is weak, the absorption
will be small unless $\Delta E_{10} \sim n\hbar\omega$.   Finally,
the integrated strength of the n-photon transition exhibits an 
conspicuous odd-even effect: the transition rate from the ground
to the first excited state generally {\em increases} with increasing $I$ for {\em even} n, but decreases for n {\em odd}.   Indeed, at zero bias current, the transition rate vanishes for even n.  This is just a consequence of the symmetry of the eigenstates in the well: if the bias current is zero, the ground and first excited states have even or odd parity, thereby allowing only odd-photon transitions.

%While our calculations do not explicitly make use of the
%anharmonicity of the washboard potential, this anharmonicity does
%play an indirect role in the calculation, because it insures that $\Delta %E_{21} \neq
%\Delta E_{10}$.  If this were not true, the microwave perturbation
%would induce transitions between the first and second excited states
%as well as between the ground and first excited states.
%IVAN2: I thought that this paragraph is unnecessary (and maybe even untrue).
%The anharmonicity is definitely included because the wave functions you
%use for states 0 and 1 are not pure harmonic oscillator states.
%DAVID2:  I agree, leave it out.

In future work, one calculation of interest would be to solve the Bloch equations
directly, i.e., without expanding the time-dependent potential
in a Bessel function series, and without making the small-amplitude
approximation which allows the absorption line to be decomposed into
a sum of Lorentzians.  Such a calculation should be straightforward to
do numerically.  Finally, of course, it would also be valuable to explicitly 
compute the additional contribution to the broadening arising from the shunt resistance in the junction.

\section{Acknowledgments}

This work was supported through NSF Grant DMR04-13395.  Calculations
were carried out, in part, using the facilities of the Ohio Supercomputer
Center with the help of a grant of time.

%DAVID, We do not use references 12 - 32 anywhere in the paper.

\newpage

\begin{tabular}{|c|c|c|c|c|c|}
\hline
i & $\phi_1$ & $\phi_2$ & $S(E_1)/\hbar$ & $\gamma_1/[\hbar\omega_0(I)]$ & $E_1/E_J$\\
\hline
0.85 & 1.2632 & 2.6636 & 6.584 & $3.04 \times 10^{-7}$ & -1.3762\\
0.90 & 1.3557 & 2.4454 & 4.740 & $1.22 \times 10^{-5}$ & -1.4336\\
0.95 & 1.5236 & 2.1451 &  2.349 & $2.35 \times 10^{-3}$ & -1.4946 \\
\hline
\end{tabular}

\vspace{0.3in}

\noindent
{\small TABLE 1: WKB action $S(E_1)/\hbar$ for three different values of the applied current
$i = I/I_c$, as indicated.  The second and third columns of the table denote the values
of the phase $\phi$ at the left and right hand edges of the barrier at energy $E_1$.  $E_1$
is the energy of the second quasi-bound state in the well, as approximated by the method
described in the text. The fourth column is the approximate ratio of the linewidth to the current-dependent
small-oscillation frequency $\omega_0(I)$. The last column gives the energy of the first excited state
$E_1(I)$, in units of $E_J$, including anharmonic corrections to lowest non-vanishing order in $\phi - \phi_0$.}

%IVAN, what are the units of $\Gamma$ here and in the other figures?
%DAVID, The units are dimensionless.
%IVAN1: Are you sure about this?  When I look at eq. (25), it seems to me the
%units would be |<0|V|1>|^2/(\hbar x E) which would have units of 1/\hbar.

%IVAN, what are the units of $\Delta E_{10}$?
%DAVID, There are no units on $\Delta E_{10}$
%IVAN1: Doesn't E_{10} have units of \hbar I_c/(2e)? 

%IVAN1: note the suggested changes mentioned in earlier captions.  Need to have conventional
%exponential notation for legends on the figure axes (e. g. 3 x 10^{-8} instead of 3e-08).
%Also, need to remove all the \Delta's from symbols like \Delta E_{10}.

\begin{figure}
\begin{center}
\includegraphics[angle=-90, width = 0.5\textwidth]{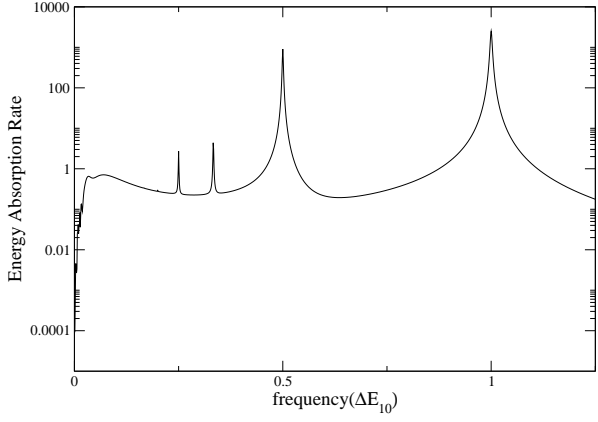}
\caption
{\small Plot of the absorption spectrum $\tilde{\Gamma}_{tot}(I, V_{ac}, \omega) = \sum_{n=1}^{nmax}\tilde{\Gamma}_n(I,V_{ac}, \omega)$ (in units of $\Delta E_{10}/\hbar$) as a function of $\omega$, for $I/I_c = 0.95$ 
and ac amplitude given by $2eV_{ac}/(\Delta E_{10}) = 0.2$, as calculated using the parameters and method described in the text.} \label{finitewidth}
\end{center}
\end{figure}

\begin{figure}
\begin{center}
\includegraphics[angle=-90, width = 0.5\textwidth]{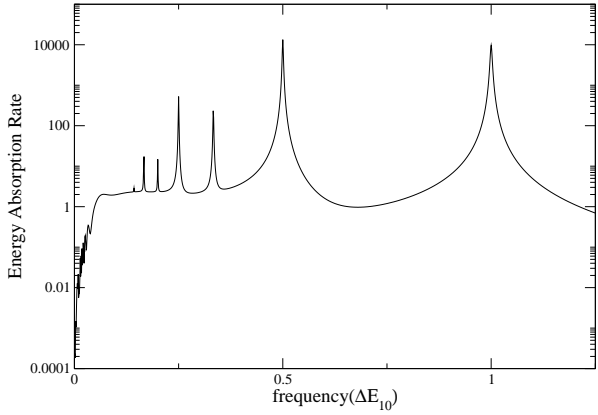}
\caption
{\small Same as Fig.\ 12, but with ac amplitude given by $2eV_{ac}/(\Delta E_{10}) = 0.4$.}\label{finitewidth1}
\end{center}
\end{figure}

\end{document}